# Mechanical Modulation of Hybrid Graphene–Microfiber Structure

Jin-hui Chen, Fei Xu* and Yan-qing Lu
*feixu@nju.edu.cn
National Laboratory of Solid State Microstructures, College of Engineering and Applied Sciences and Collaborative Innovation Center of Advanced Microstructures, Nanjing University, Nanjing 210093, P. R. China

**Abstract**

Recently, the strain engineering of two-dimensional materials such as graphene has attracted considerable attention for its great potential in functional nanodevices. Here, we theoretically and experimentally investigate the strain manipulation of a graphene-integrated microfiber system for the first time. We analyze the influential factors of strain tuning, i.e., the geometrical parameters of the microfiber, the strain magnitude, and the probe-light wavelength. Moreover, we experimentally achieve in-line modulation as high as 30% with a moderate strain of ~5%, which is two orders of magnitude larger than previous results. The dynamic vibration response is also researched. The broadband, polarization-independent, cost-effective, strain-based modulator may find applications in low-speed modulation and strain sensing. Further, we believe that our platform may allow for all-in fiber engineering of graphene-analogue materials and provide new ideas for graphene-integrated flexible device design.

**Keywords**： graphene, polarization, microfiber, strain, modulation

In the past ten years, there have been many breakthroughs in graphene research ranging from massless Dirac fermions, the half-integer quantum Hall effect [1], and the Klein paradox [2] to the fine-structure constant [3], etc. Many novel applications based on graphene have been proposed, such as optical modulators [4], photodetectors [5-7], mode-locked lasers [8], and molecular sensors [9]. In contrast to its three-dimensional (3D) counterpart, graphene has a two-dimensional (2D) nature and is very sensitive to changes in the ambient environment. Further, many techniques have been developed to manipulate graphene's electronic and optical properties, from electric gating [4, 10] and chemical doping [9] to strain engineering [11-14].

Strain engineering has been recognized as an effective way to tune the properties of low-dimensional materials and to enhance the performance of nanodevices [12, 14-18]. For one-dimensional (1D) nanomaterials, many experimental works have employed piezoelectric nanowires to harvest mechanical energy from the ambient environment for self-powered devices [15, 16]. Recently, the increasing development of two-dimensional materials, i.e., graphene and graphene-analogue materials, have provided more possibilities for manipulation. For transition-metal dichacolgenides (TMDs) such as $MoS_2$, the inherent centrosymmetry broken for odd number of layers creates giant piezoelectricity, which may enable adaptive bioprobes and nanoelectromechanical systems [17, 18]. For graphene, it was predicted and experimentally demonstrated that lattice distortions of graphene can generate a pseudomagnetic field as high as 300 T, which leads to the pseudo-quantum Hall effect [13, [14]. Moreover, it was found that the Raman fingerprints of graphene are shifted and split under strain [11]. In contrast to the electronic properties, there are only a few works that report the tuning of graphene's optical properties by strain [19-21]. Although graphene has an impressive 2.3% absorption per layer in the visible and near-infrared spectra [3, 22], the anisotropic-polarization-dependent absorption created by a uniaxial strain [20] is still negligible, thus limiting its further applications. Recently, a waveguide-integrated-graphene platform [4, 7, 23-25] was demonstrated to be effective for increasing the light–graphene interaction length (strength) and improving the device's functionality. In addition, microfibers (MFs) have received special attention [26, 27] for their strong confinement, large evanescent fields, great mechanical strength, and low-loss connection.

Here, we theoretically and experimentally investigate the strain manipulation of a hybrid graphene–integrated microfiber (GMF) structure for the first time. By in-line stretching of an MF coated with a piece of graphene, the strain in the MF will induce a uniaxial strain in the graphene, which leads to a change in the waveguide absorption loss [19, 20]. According to our theoretical calculations, we find that the decrease in the MF volume also contributes to the modulation of the GMF because of the enhanced evanescent field in the MF. Owing to the ultralong light–graphene interaction length (tens of millimeters), we have achieved ~30% polarization-independent modulation with a moderate strain of ~5%, which is two orders magnitude larger than previous results [20]. We also characterize the dynamic mechanical response of the GMF from 50 Hz to 1 kHz. We believe our platform may allow for all-in fiber engineering of graphene and provide applications in graphene-integrated flexible devices and strain sensing.

**Results**

**Characterization of the GMF.** Figure 1a shows a schematic of the structure of the GMF, in which a piece of graphene is circularly wrapped around the waist of an MF. To fabricate the GMF, an MF with a diameter of 5–10 μm was first tapered from a standard optical fiber (SMF-28, Corning, New York, USA). Then, the graphene on a copper foil (Six Carbon Technology, Shenzhen, China) was etched with a 1 M $FeCl_3$ aqueous solution and rinsed in deionized water several times. After that, the MF was dip-coated with graphene (**Supplementary Figure S1, Supplementary Note 1**) that was several millimeters long. Figure 1b shows the as-fabricated GMF fixed on a translational stage for further strain response measurement. The inset shows a camera image of the graphene-deposited areas on the GMF. The strong scattering of red light is caused by the graphene coating [23]. The scanning electron microscopy (SEM) image of the graphene morphology on the MF in Fig. 1c indicates a considerably good surface topography, although there are some wrinkles and contaminations on the graphene, which may be introduced during the transfer process. Raman spectroscopy is utilized to qualitatively characterize the quality of the transferred graphene, as is shown in Fig. 1d. The two most intense features are the 2D peak at 2658 $cm^{-1}$ and the G peak at ~1584 $cm^{-1}$, and their intensity contrast verifies the monolayer nature of graphene [28]. The D peak at 1330 $cm^{-1}$ is also observed, which indicates defects in graphene.

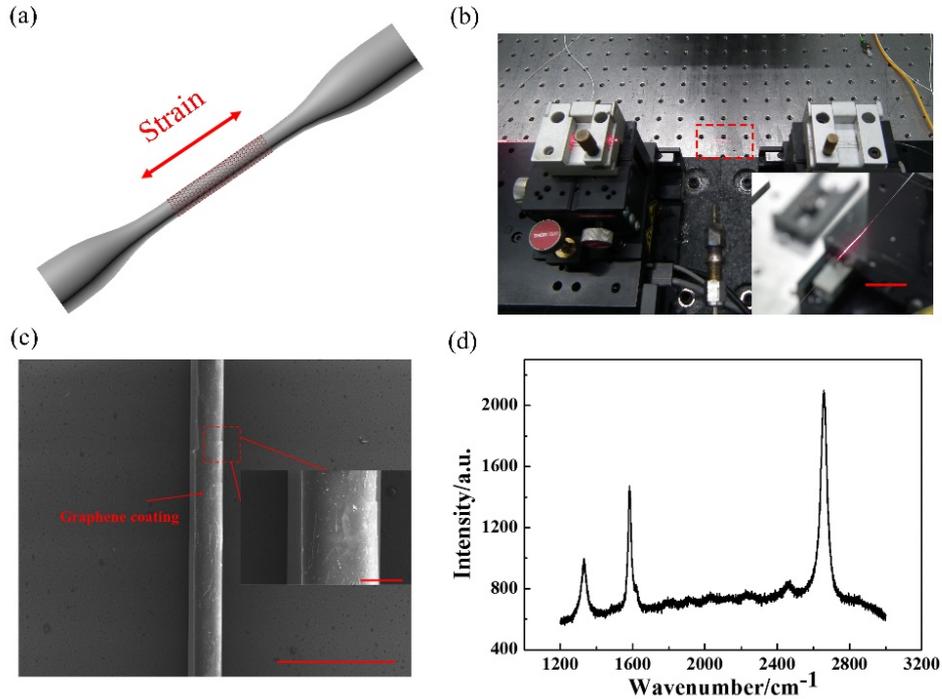

**Figure 1 | Characterization of the GMF structure.** (**a**) Schematic of the GMF structure. Part of an MF surface was coated with graphene. (**b**) Camera image of the fabricated GMF fixed on a translational stage for strain measurement. The inset shows a magnification of the GMF in the red-dash box. The GMF is launched with a red-light source, and the strong scattering area indicates the deposited graphene on the MF. The scale bar is 1 cm. (**c**) SEM image of the as-fabricated GMF. The red arrow indicates deposited graphene. The scale is 40 μm. The inset shows the amplified image obtained from the dashed box, and the scale is 5 μm. (**d**) Raman spectra of graphene on the MF.

**Strain-induced modulation of the GMF.** To achieve in-line manipulation of the GMF, we first fixed the as-fabricated GMF on a translational stage, as illustrated in Fig. 1b. With a computer-assisted monitoring system, we can precisely control the strain in the GMF. Figure 2 shows the transmission modulation of the GMF induced by strain. We used a nonpolarized light source input, i.e., an amplified spontaneous emission (ASE) light source (Connet, C-ASE Optical Source), which covers spectra of 1530–1565 nm. The diameter of the GMF is ~6 μm, and the length of the deposited graphene is ~8 mm. As seen in Fig. 2a, the transmission of the GMF decreases as the strain increases. Here, we define the modulation as the change in the transmission of the GMF. Figure 2b clearly shows that the modulation of the GMF is almost linear with strain, and the maximum value is ~1.1 dB (equal to ~30% modulation). Moreover, the modulation also weakly depends on the light wavelength, as illustrated in Fig. 2c. It should be noted that the GMF's transmission has only a small dependence on the input light polarization theoretically and experimentally (**Supplementary Figure S2**) because of the circular symmetry of the GMF.

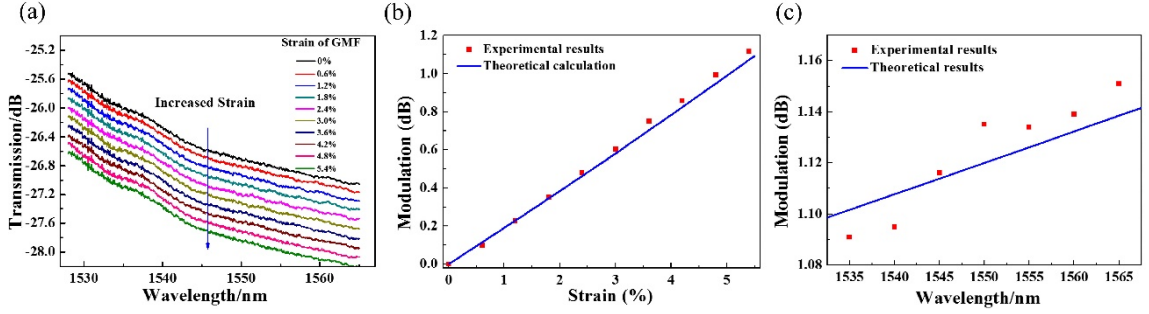

**Figure 2 | Strain modulation of the GMF transmission.** (**a**) Change in the transmission of the ASE light source for various strains. (**b**) Modulation depth of the GMF with a uniaxial strain at 1545 nm, which is extracted from (**a**). The blue line denotes the theoretical results. (**c**) Wavelength-dependent modulation of the GMF caused by 5.4% strain, which is retrieved from (**a**). The blue line denotes the theoretical results.

We developed a theoretical model [29] (**Supplementary Note 3**) to explain the observed phenomenon. It is well known that graphene's optical conductivity will become anisotropic under a uniaxial strain [20, 21]:

$$\sigma_{T,L}(\omega) \approx \sigma_0(\omega)[1 \pm p(1+v)\kappa] \qquad (1)$$

where $\sigma_T$ ($\sigma_L$) represents the optical conductivity perpendicular to the direction of the strain (parallel to the direction of the strain), $\sigma_0$ is the optical conductivity in the strain-free case, $p$ is a numerical constant and $\approx$3–4 [21], and $v$ is the Poisson ratio. From (1), we can determine when the graphene sheet is under a uniaxial strain; its optical conductivity along the direction of the strain decreases while the transverse conductivity increases. Thus, when light nominally illuminates strained graphene, the light polarization transverse to the direction of the strain will suffer a larger absorption loss than the parallel one [20], which leads to anisotropic polarization absorption. However, things will be much different for the waveguide case. First, most of the electric field (magnetic field) of the eigenmodes in the microfiber waveguide lie in the transverse plane. Thus, the longitudinal conductivity of graphene (in cylindrical coordinates) (**Supplementary Note 3**) has a smaller influence than the transverse optical conductivity on the waveguide's absorption loss caused by graphene. As a result, when the GMF is elongated in the axial direction, the strain-induced modulation of the waveguide's transmission is dominated by the change in the transverse optical conductivity. As the transverse conductivity of graphene increases with the uniaxial strain, the waveguide transmission will naturally decrease with the increase in the strain, which is clearly illustrated in Fig. 2. Second, thanks to the symmetry of the GMF, there theoretically will be no polarization-dependent absorption effect for the wave-guiding modes. However, the fabricated device has a smaller polarization-dependent loss because of the defects in the sample. Interestingly, according to our theory, we find that the decrease in the volume of the GMF caused by strain also contributes to the modulation phenomenon (**Supplementary Note 3**), and its influence on the modulation of the GMF is comparable to the change in the optical conductivity of graphene. This is because the waveguide's propagation loss is highly sensitive to this geometrical change, as shown in Fig. 3a. The wavelength-dependent modulation relation can be attributed to the fact that the evanescent field strength of the MF has a weak wavelength dependence; that is, the longer light wavelength of the GMF has a larger evanescent field to interact with the graphene coatings, and thus, a larger absorption loss (Fig. 2a) and larger modulation (Fig. 2c) are obtained. Our theoretical calculations agree well with the experimental data.

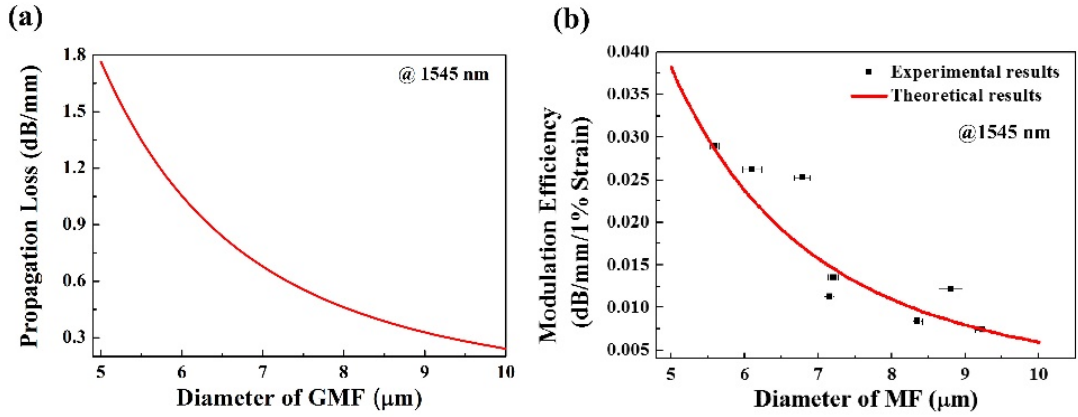

**Figure 3 | Strain-induced modulation relations with the diameter of the GMF.** (a) Propagation loss of the waveguide as a function of the diameter of the GMF at 1545 nm. (b) Strain modulation efficiency of the GMF as a function of the diameter of the MF at 1545 nm.

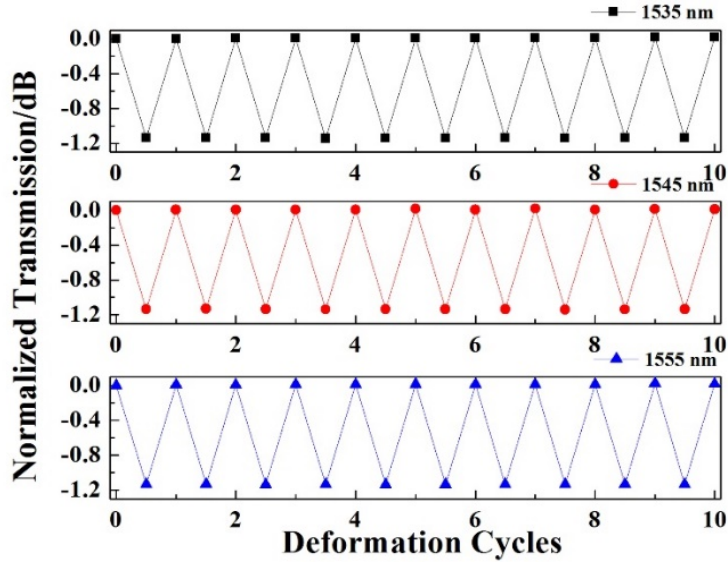

**Figure 4 | Stable deformation cycling test of the GMF at 1535 nm, 1545 nm, and 1555 nm.**

We also investigate the influence of the MF diameter on the strain-induced modulation of the GMF. Here, we define the modulation efficiency (ME) as the modulation per unit length of graphene-coated MF per 1% strain to quantitatively assess the impact of the MF diameter. It is clearly shown that the ME of the GMF exponentially decreases as the diameter increases in Fig. 3a. As is known, the evanescent field strength decreases as the waveguide diameter increases. Accordingly, the light–graphene interaction strength decreases as the diameter of the waveguide increases, and the ME of the strain is also naturally decreases. The stability and repeatability of our platform are also characterized, as illustrated in Fig. 4. It is demonstrated that the strain cycling is repeatable and stable, which verifies that our device is elastically deformed. Further, the modulation effect has a weaker dependence on the light wavelength, as explicitly discussed above. To test the GMF's dynamic strain response, we employed dynamic vibration of the GMF (**Supplementary Note 4**), as shown in Fig. 5. Although the vibration frequencies only cover 50–1000 Hz, we believe that it can achieve a modulation speed of hundreds of kilohertz by further careful design. The broadband, polarization-independent, cost-effective strain modulator may find applications in low-speed modulation, strain sensing, and waveguide design.

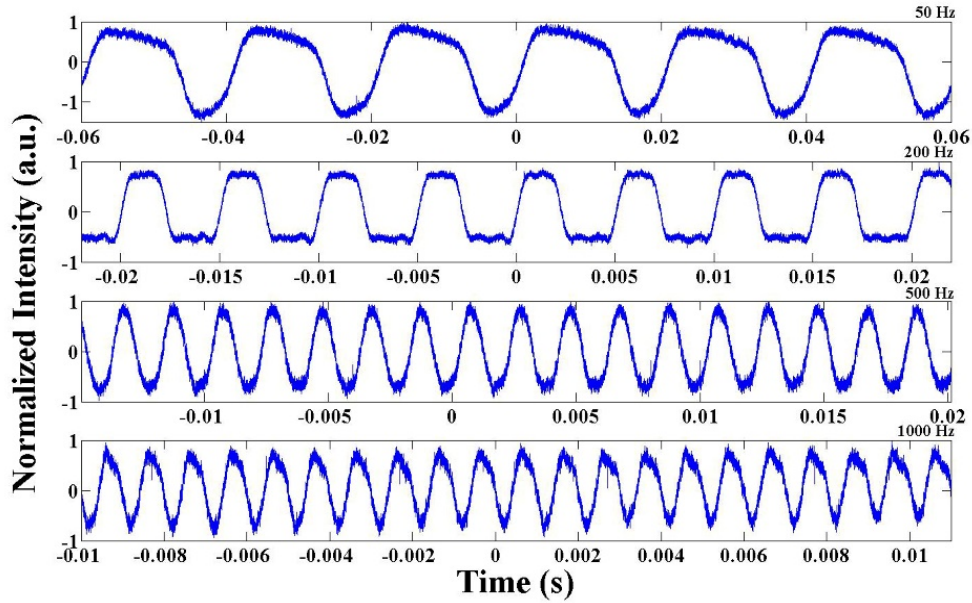

**Figure 5 | Dynamic mechanical-vibration-induced modulation of the GMF.** The mechanical vibration response was tested at 50 Hz, 200 Hz, 500 Hz, and 1 kHz separately.

**Discussion**

It has been proposed that the uniaxial-strain manipulation of a graphene sheet can lead to the anisotropic optical conductivity of graphene. Here, we investigate waveguide-based amplification due to the strain effect in graphene for the first time. We theoretically analyze the influential factors of strain tuning, such as the geometrical parameters of the MF, the magnitude of the strain, and the light wavelength. Further, we experimentally achieve modulation as high as 30% with a moderate strain of ~5%, which is two orders of magnitude larger than previous results. Moreover, the modulation can be further improved, e.g., by tuning the encapsulated graphene length, the diameter of the MF, and the employed strain. It should be noted that graphene can be elastically deformed to a strain as high as 25% [30], which indicates the great potential of strain manipulation for waveguide-integrated graphene. We believe our platform may allow for the all-in fiber engineering of graphene and provide new ideas for graphene-integrated flexible devices and strain sensing.

**Methods**
**SEM and Raman measurements**
The surface morphology of the graphene transferred onto an MF was characterized by a dual beam focus ion beam system (FIB) 235, FEI Strata. Raman spectra were obtained by a Horiba John Yvon HR800 system with a laser excitation wavelength of 633 nm.
**Strain response measurement**
The in-line stretching of the GMF was generated by a linear motor stage with a travel distance of 350 mm (XML, Newport). The dynamic vibration was produced by a homemade speaker with its output controlled by a program.

**References**

1. Novoselov K, *et al.* Two-dimensional gas of massless Dirac fermions in graphene. *Nature* **438**, 197-200 (2005).
2. Katsnelson M, Novoselov K, Geim A. Chiral tunnelling and the Klein paradox in graphene. *Nat. Phys.* **2**, 620-625 (2006).
3. Nair RR, *et al.* Fine structure constant defines visual transparency of graphene. *Science* **320**, 1308-1308 (2008).
4. Liu M, *et al.* A graphene-based broadband optical modulator. *Nature* **474**, 64-67 (2011).
5. Xia F, Mueller T, Lin Y-m, Valdes-Garcia A, Avouris P. Ultrafast graphene photodetector. *Nat. Nanotechnol.* **4**, 839-843 (2009).
6. Mueller T, Xia F, Avouris P. Graphene photodetectors for high-speed optical communications. *Nat. Photonics* **4**, 297-301 (2010).
7. Gan X, *et al.* Chip-integrated ultrafast graphene photodetector with high responsivity. *Nat. Photonics* **7**, 883-887 (2013).
8. Sun Z, *et al.* Graphene mode-locked ultrafast laser. *ACS Nano* **4**, 803-810 (2010).
9. Schedin F, *et al.* Detection of individual gas molecules adsorbed on graphene. *Nat. Mater.* **6**, 652-655 (2007).
10. Wang F, *et al.* Gate-variable optical transitions in graphene. *Science* **320**, 206-209 (2008).
11. Ni ZH, Yu T, Lu YH, Wang YY, Feng YP, Shen ZX. Uniaxial strain on graphene: Raman spectroscopy study and band-gap opening. *ACS Nano* **2**, 2301-2305 (2008).
12. Pereira VM, Neto AC. Strain engineering of graphene's electronic structure. *Phys. Rev. Lett.* **103**, 046801 (2009).
13. Guinea F, Katsnelson M, Geim A. Energy gaps and a zero-field quantum Hall effect in graphene by strain engineering. *Nat. Phys.* **6**, 30-33 (2010).
14. Levy N, *et al.* Strain-induced pseudo–magnetic fields greater than 300 tesla in graphene nanobubbles. *Science* **329**, 544-547 (2010).
15. Wang ZL, Song J. Piezoelectric nanogenerators based on zinc oxide nanowire arrays. *Science* **312**, 242-246 (2006).
16. Xu S, Qin Y, Xu C, Wei Y, Yang R, Wang ZL. Self-powered nanowire devices. *Nat. Nanotechnol.* **5**, 366-373 (2010).
17. Zhu H, *et al.* Observation of piezoelectricity in free-standing monolayer $MoS_2$. *Nat. Nanotechnol.* **10**, 151-155 (2015).
18. Wu W, *et al.* Piezoelectricity of single-atomic-layer $MoS2$ for energy conversion and piezotronics. *Nature* **514**, 470-474 (2014).
19. Pellegrino F, Angilella G, Pucci R. Strain effect on the optical conductivity of graphene. *Phys. Rev. B* **81**, 035411 (2010).
20. Ni GX, *et al.* Tuning Optical Conductivity of Large‐Scale CVD Graphene by Strain Engineering. *Adv. Mater.* **26**, 1081-1086 (2014).
21. Pereira VM, Ribeiro R, Peres N, Neto AC. Optical properties of strained graphene. *EPL (Europhysics Letters)* **92**, 67001 (2010).
22. Mak KF, Sfeir MY, Wu Y, Lui CH, Misewich JA, Heinz TF. Measurement of the optical conductivity of graphene. *Phys. Rev. Lett.* **101**, 196405 (2008).
23. Li W, *et al.* Ultrafast all-optical graphene modulator. *Nano Lett.* **14**, 955-959 (2014).

24. Kou J-l, Chen J-h, Chen Y, Xu F, Lu Y-q. Platform for enhanced light–graphene interaction length and miniaturizing fiber stereo devices. *Optica* **1**, 307-310 (2014).
25. Bao Q*, et al.* Broadband graphene polarizer. *Nat. Photonics* **5**, 411-415 (2011).
26. Tong L*, et al.* Subwavelength-diameter silica wires for low-loss optical wave guiding. *Nature* **426**, 816-819 (2003).
27. Brambilla G*, et al.* Optical fiber nanowires and microwires: fabrication and applications. *Adv. Opt. Photonics* **1**, 107-161 (2009).
28. Ferrari A*, et al.* Raman spectrum of graphene and graphene layers. *Phys. Rev. Lett.* **97**, 187401 (2006).
29. Gao Y, Ren G, Zhu B, Liu H, Lian Y, Jian S. Analytical model for plasmon modes in graphene-coated nanowire. *Opt. Express* **22**, 24322-24331 (2014).
30. Kim KS*, et al.* Large-scale pattern growth of graphene films for stretchable transparent electrodes. *Nature* **457**, 706-710 (2009).

**Acknowledgements**

The author thanks Yue-hui Chen for helping Raman characterizations. The SEM was supported by Ying-ling Tan. This work is sponsored by National 973 program (2012CB921803), and National Natural Science Foundation of China (61322503, 61225026, and 61475069).